\documentstyle[12pt,titlepage]{article}

\newcommand{\gorder}{$\stackrel{>}{\sim}$}
\newcommand {\h} {$h^{-1} \, Mpc \,$}
\def\ref{\par\noindent\hangindent=1cm}
\begin{document}
\title{}
\author{}
\date{}
\vspace{4cm}
\Huge
\begin{center}
{\bf The Mass Function \protect\\
of Nearby Galaxy Clusters}
\end{center}
\vspace{2cm}
\normalsize
\begin{center}
{\em A. Biviano$^{1,2}$, M. Girardi$^{1}$, \protect\\
G. Giuricin$^{1,3}$, F. Mardirossian$^{1,3}$, and M. Mezzetti$^{1,3}$}
\end{center}
\vspace{0.5cm}
(1) SISSA, strada Costiera 11, 34014 -- Trieste, Italy \\
(2) Institut d'Astrophysique de Paris, 98bis, bd. Arago, 75014 Paris, France \\
(3) Dipartimento di Astronomia,
Universit\`{a} degli Studi di Trieste, Italy \\
\pagebreak
\renewcommand{\thesection}{\arabic{section}}
\renewcommand{\thesubsection}{\thesection.\arabic{subsection}}
\baselineskip 28pt
\section*{Abstract}
We present the distribution of virial masses for nearby
galaxy clusters, as obtained from a data-set of 75 clusters,
each having at least 20 galaxy members with measured
redshifts within 1.5 \h. After having accounted for problems
of incompleteness of the data-set, we fitted a power-law to
the cluster mass distribution.
\\
\vspace{1cm}
{\em Subject headings:} \ galaxies: clustering
\pagebreak
\section{Introduction}
The distribution functions of observational cluster quantities, such as
velocity dispersions, radii, masses, luminosities and X-ray temperatures,
can provide strong constraints both on cosmological scenarios and on
the internal dynamics of these systems. Theoretical as well as
observational estimates of these distribution functions are presently
being debated.

The theoretical mass distribution expected for groups and clusters in the
hierarchical clustering scenario has been derived by several authors
(see, e.g., Press \& Schechter 1974; Cavaliere, Colafrancesco, \& Scaramella
1991; Blanchard, Valls-Gabaud, \& Mamon 1992). The distribution functions
for mass, X-ray temperature and velocity dispersion of rich clusters have been
shown to provide important diagnostics for cosmological models (see, e.g.,
Edge et al. 1990, Henry \& Arnaud 1991, Lilje 1992).
In particular,
the distribution function of cluster X-ray temperatures has been used
to constrain the mass fluctuation spectrum (see, e.g., Henry \& Arnaud 1991).
Bahcall (1979) estimated the cluster luminosity function,
and first offered the possibility of
evaluating the cluster mass function by adopting a constant mass-to-light
ratio. At present both optical and X-ray data contribute to
the description of the cluster mass function (Bahcall \& Cen 1992).
Recently, Pisani et al. (1992) obtained
the mass distribution function of groups of galaxies directly from
their dynamics, while Ashman, Persic, \& Salucci (1993) described
a distribution function of galaxy masses.

Quite a large number of redshifts for cluster galaxies is
now available, so that many cluster masses can be measured
directly from dynamics. We collected from the literature 75
clusters, each having at least 20 members with measured redshift
within 1.5 \h (we use $H_{0}= 100 \, h \, km \, s^{-1} \, Mpc^{-1}$
throughout). In Table 1 we list the clusters considered, the number $N$
of cluster members with available redshift, and the richness classes $R$
mainly taken from Abell, Corwin, \& Olowin (1989; ACO hereafter).

In order to reasonably reduce evolutionary
effects, we considered only clusters with mean redshift $z \leq 0.15$.
In Biviano et al. (1992), Girardi et al. (1993a),
and Girardi et al. (1993b) we describe the data-base and we give the relevant
references. In Girardi et al. (1993a) we detail the criteria
used to define the clusters both in redshift space and in projected
separation.

We have shown (Girardi et al. 1993a, and 1993b)
that at least 20 galaxy redshifts are generally
needed to describe cluster parameters (in particular the galaxy velocity
dispersion and the virial radius) without heavy problems induced by
undersampling. So, from our statistical point of view,
the choice of a minimum of 20 redshift per cluster seems to be
an acceptable compromise between the need of having at disposal
a large number of clusters and the possible presence of undersampling problems.

Girardi et al. (1993a) studied, in a homogeneous way, the
velocity field of galaxy clusters. They evidenced that
the distribution of galaxy velocities in
clusters is quasi-Gaussian, which encouraged us to apply the
Virial Theorem to estimate the cluster masses. Moreover,
the estimates of cluster masses, via the Virial Theorem, have
been widely debated and accepted in recent literature.
As shown by, e.g., Postman, Geller, \& Huchra (1988),
Perea, del Olmo, \& Moles (1990), and
Pisani et al. (1992), the Virial Theorem measures
masses with confidence comparable to that of other methods,
in particular the Projected Mass estimator
(see Heisler, Tremaine, \& Bahcall, 1985).

Our virial masses have been obtained by using
robust estimates of the velocity dispersions
(Beers, Flynn, \& Gebhardt 1990, Girardi et al. 1993a).
The mass errors quoted are obtained via the jackknife
method (see, e.g., Efron 1979, and Geller 1984,
for an astrophysical application). In the present paper we
evaluate the virial masses both using two apertures, 0.75 \h
and
1.5 \h, i.e., half and one Abell radius, respectively.
We also computed the projected
masses of our clusters, within the same aperture.
Remarkably, the shape of the distributions
of virial and projected masses is almost the same.

The presence of subclustering may bias cluster mass estimates.
In particular the bias may be severe, and the mass may be
overestimated,  when the subclustering
masks the presence of two or more unbound systems. In this
case one should consider the clumps as isolated clusters or
groups. We adopted the method of Dressler \& Shectman (1988) to
investigate the presence of subclustering in our sample.
Actually this method (like other methods present
in the literature) is efficient for a number of members
$N$ \gorder 40; for a lower number of members the efficiency decreases;
therefore we considered only well sampled clusters with $N \geq 40$.
The probability of subclustering was computed
using all the galaxies within 1.5 \h
 from the cluster center;
a probability $\geq$ 95 \% was taken to be significant.

The mass distribution of the 18
clusters without evidence for subclustering  did not succeed to be
significantly
different (according to the Mann-Whitney U-test, see,
e.g., Siegel 1956) from that of the 12 clusters
with evidence of subclustering.
We obtained again the same result when we compared
the distributions of masses computed inside 0.75 \h
 (27 vs. 16 clusters).
This result suggests that the effect of subclustering does not strongly
affect our mass distribution, either because it is small, or because our
selection criteria for cluster membership and our
estimate of the velocity dispersions are statistically robust (see
Girardi et al. 1993a). Therefore, no cluster has been rejected.
In view of possible surviving cases in which subclustering effects produce
mass-overestimates, it is conservative to consider mass estimates
obtained within small apertures. As a matter of
fact,  West \& Bothun (1990) evidenced that only a few significant
subclumps exist within 1 \h  in $\sim 70$ clusters analyzed.

Of course, other systematic errors in the mass evaluation
may be present if light does not trace mass in our clusters
(see, e.g., Merritt 1987, Watt et al. 1992; and references therein).

Masses are in solar units.

\section{The Distribution of Cluster Masses}
In the above-mentioned Table 1 we also list the logarithms of the cluster
virial
masses $\log (h M_{0.75})$, and $\log (h M_{1.5})$ measured within apertures of
0.75 \h, and 1.5 \h.
In order to measure mass at a certain radius, we must have information
on the galaxy velocity-field and the gravitational potential
up to the same radius. Therefore, we have estimated $M_{0.75}$
only for those 69 clusters which are actually well sampled
up to 0.75 \h from the cluster center;
these clusters contain $\geq 20$ galaxies all located
within 0.75 \h, and the most external of these galaxies is
located, in each cluster, at $\simeq 0.75$ \h from the center.
Similarly, we have estimated
$M_{1.5}$ only for those 47 clusters which are actually well
sampled up to 1.5 \h from the cluster center;
these clusters contain $\geq 20$ galaxies all located
within 1.5 \h, and the most external of these galaxies is
located, in each cluster, at $\simeq 1.5$ \h from the cluster center.

The masses $M_{0.75}$ and $M_{1.5}$ both
contribute to describe the cluster mass distribution.
In fact, fixed apertures may contain different fractions of cluster
masses (depending on the intrinsic cluster sizes), and
define two different cuts of the density
peaks which identify the clusters in the cosmic density field.
However, in the present paper, we restrict our analysis
to the galaxies contained within 0.75 \h, mainly
because of the above-mentioned problem of subclustering.
The distribution of $M_{0.75}$ is plotted in Fig.1.

Our sample, although quite extended, is not complete, either in
richness or in mass. Therefore, we
normalized the frequencies of clusters belonging to different
richness classes to the corresponding frequencies of ACO's Catalogue;
the normalization procedure is described in
Girardi et al. (1993a). However, since
the completeness of ACO's Catalogue
for $R=0$ clusters is in question (see, e.g., Scaramella et al. 1991, and
Guzzo et al. 1992), we also considered the Edinburgh-Durham Cluster
Catalogue (EDCC, hereafter) of Lumsden et al. (1992), by
re-scaling their richness to ACO's.
Moreover, using two catalogues also allows us to investigate
the sensitivity of the results on the choice of the catalogue.
In Fig.2, we show the histograms of the
ACO-normalized and EDCC-normalized
mass distributions of our clusters.

\section{The Mass Function}
As one can see in Fig.2, the mass distributions do not increase
monotonically from high to low mass values.
Two facts may contribute to the decrease at low mass values: (i)
the possible incompleteness of ACO's and the Edimburgh-Durham
Cluster Catalogues for low richness
clusters; (ii) the lack, in our sample, of rich galaxy groups, the masses
of which are expected to fall partially in the low-mass region (in fact,
completeness in richness does not necessarily imply completeness in mass).

The problem of the non-monotonic behaviour of the mass distribution will
not be solved unambiguously until a complete sample of galaxy systems,
spanning the range from small groups to rich clusters, is available. So,
even if more data on poor cluster and rich groups (which are becoming
available)
were considered, we would
still deal with an incomplete sample, and the re-normalization problem
would not be solved.

So we deemed it conservative to use only the high-mass (decreasing)
part of the distribution in order
to obtain the mass function of nearby clusters. Specifically, we set
a lower bound of $\log (h M_{0.75}) = 14.6$
to the mass distributions.

We fitted these two culled distributions with power laws,
$n(M_{15}) dM_{15} = A (M_{15})^a dM_{15}$, where $M_{15}$ is the mass
expressed in units of $10^{15} M_{\odot}$.
These power laws were convolved with Lognormal functions
whose dispersions were obtained from
the median values of the cluster mass errors (quoted in
Table 1). The best-fit values of $a$ (obtained via the
Maximum Likelyhood method) are
$-2.3^{+0.6}_{-0.6}$ and $-2.6^{+0.7}_{-0.6}$
for the ACO-normalized and EDCC-normalized distributions,
respectively. These $a$ values are not different within the errors,
and the result does not seem to be very sensitive on the choice of the
reference catalogue.

The volume density of ACO clusters has been estimated to be
$2.5 \times 10^{-5} h^{3} Mpc^{-3}$ for $R \geq 0$ by
Scaramella et al. 1991, and Zucca et al. 1993, and to be
$8.6 \times 10^{-6} h^{3} Mpc^{-3}$ and
$7.2 \times 10^{-6} h^{3} Mpc^{-3}$ for $R = 0$ and $R \geq 1$,
respectively, by Peacock \& West (1992). Thus
the mass function, in the assumed range of completeness, is given by
$n(M_{15}) = 5 \times 10^{-6} h^{1.7}
(M_{15})^{-2.3} Mpc^{-3} (10^{15} M_{\odot})^{-1}$,
if one considers the density estimate of Scaramella et al. (1992) and
Zucca et al. (1993),
or $n(M_{15}) = 3.2 \times 10^{-6} h^{1.7}
(M_{15})^{-2.3} Mpc^{-3} (10^{15} M_{\odot})^{-1}$,
if one considers the density estimate of Peacock \& West (1992).

It is also possible to give an estimate of the exponent $n$ of the power law
spectrum $P(k) \propto k^n$ of cosmic fluctuations. A
way to do that is to compare the slope
of the mass function with that one of the power law representation of the
temperature function of clusters given by Henry \& Arnaud (1991) and
Henry (1992). By making use of both of these functions we
can better constrain the value of $n$ than relying only on
the mass function, which admittedly span a short range in cluster masses.
In the framework of self-similar clustering (Kaiser 1990),
virial mass $M$ and temperature $T$ are linked by $T \propto M^{(1-n)/6}$.
Comparing the slopes of the temperature function ($-4.75 \pm 0.50$)
and of the mass functions gives a spectral index
$n = -1.1^{+0.9}_{-0.9}$ and $n = -1.6^{+1.1}_{-1.0}$
for ACO and EDCC normalized distributions, respectively.

In conclusion, we have presented a cluster mass function, obtained directly
form the dynamics of these galaxy systems, which is described by
a power law, and which
mainly refers to high mass clusters. In the future we plan to extend
this mass function to a larger mass range, including dynamical estimates of
masses of poor clusters and groups of galaxies.

\vspace{2cm}

We acknowledge useful discussions with
Alfonso Cavaliere, Francesco Luc\-chin, Massimo Persic, Armando Pisani, and
Paolo Salucci. We thank an anonymous Referee for his/her suggestions.
We wish to thank Luigi Guzzo for kindly
providing us with a digitized version of the EDCC, and Dario Fadda
for his help in the data-set compilation.
This work was partially supported by the {\em
Ministero per l'Universit\`a e per la Ricerca scientifica e tecnologica},
and by the {\em Consiglio Nazionale delle Ricerche (CNR-GNA)}.

\pagebreak
\section*{References}

\ref Abell, G.O., Corwin, H.G.Jr., \& Olowin, R.P. 1989, ApJS, 70, 1 (ACO).

\ref Ashman, K.M., Salucci, P., \& Persic, M., 1993, MNRAS, 260, 610.

\ref Bahcall, N.A. 1979, ApJ, 232, 689.

\ref Bahcall, N.A., \& Cen, R. 1992, ApJ, 398, L81.

\ref Beers, T.C., Flynn, K., \& Gebhardt, K. 1990,  AJ,  100, 32.

\ref Biviano, A., Girardi, M., Giuricin, G., Mardirossian, F., \& Mezzetti, M.
1992, ApJ, 376, 458.

\ref Blanchard, A., Valls-Gabaud, D., \& Mamon, G.A. 1992, A\&A, 264, 365.

\ref Cavaliere, A., Colafrancesco, S., \& Scaramella, R. 1991, ApJ, 380, 15.

\ref Dressler, A., \& Shectman, S.A. 1988, AJ, 95, 985.

\ref Edge, A.C., Stewart, G.C., Fabian, A.C., \& Arnaud, K.A. 1990, MNRAS,
245, 559.

\ref Efron, B., 1979, S.I.A.M. Rev, 21, 460.

\ref Geller, M.J. 1984, in {\em Clusters and Groups of Galaxies},
F. Mardirossian, G. Giuricin, \& M.Mezzetti Eds., D. Reidel Pub. Com.
,Dordrecht, Holland, p. 353.

\ref Girardi, M., Biviano, A., Giuricin, G., Mardirossian, F., \&
Mezzetti, M. 1993a, ApJ in press.

\ref Girardi, M., Biviano, A., Giuricin, G., Mardirossian, F., \&
Mezzetti, M. 1993b, in preparation.

\ref Guzzo, L., Collins, C.A., Nichol, R.C., \& Lumsden, S.L. 1992, ApJ, 393,
L5.

\ref Heisler, J., Tremaine, S., \& Bahcall, J.N., 1985, ApJ, 298, 8.

\ref Henry, J.P., \& Arnaud, K.A. 1991, ApJ, 372, 410.

\ref Henry, J.P. 1992, in {\em Clusters and Superclusters of Galaxies},
A.C. Fabian Ed., Kluwer Academic Publishers, Dordrecht, Holland, p. 311.

\ref Kaiser, N. 1990, in {\em Clusters of Galaxies}, W.R. Oegerle, M.J.
Fitchett, \& L. Danly (Eds.), Cambridge University Press, Cambridge, U.K.,
p. 327.

\ref Lilje, P.B. 1992, ApJ, 386, L33.

\ref Lumsden, S.L., Nichol, R.C., Collins, C.A., \& Guzzo, L. 1992, MNRAS,
258, 1.

\ref Merritt, D., 1987, ApJ, 313, 121.

\ref Peacock, J.A., \& West, M.J. 1992. MNRAS 259, 494.

\ref Perea, J., Del Olmo, A., \& Moles, M. 1990, A\&A, 237, 319.

\ref Pisani, A., Giuricin, G., Mardirossian, F., \& Mezzetti, M.
1992, ApJ, 389, 68.

\ref Postman, M., Geller, M.J., \& Huchra, J.P 1988, AJ, 95, 267.

\ref Press, W.H., \& Schechter, P. 1974, ApJ, 187, 425.

\ref Scaramella, R., Zamorani, G., Vettolani, G., \& Chincarini, G. 1991,
AJ, 101, 342.

\ref Siegel, S., 1956, {\em Non Parametric Statistics for the Behavioral
Sciences}, McGraw-Hill Book Co., New York, p.68.

\ref Watt, M.P., Ponman, T.J., Bertram, D., Eyles, C.J., Skinner, G.K.,
\& Willmore, A.P. 1992, MNRAS, 258, 738.

\ref West, M.P., \& Bothun, G.D, ApJ, 350, 36.

\ref Zucca, E., Zamorani, G., Scaramella, R., \& Vettolani, G. 1993,
ApJ, in press.

\pagebreak
\section*{Captions to Figures}
{\bf Fig.1:} The cumulative distributions of
$\log (h M_{0.75})$
for our 69 clusters with data up to 0.75 \h.

\noindent {\bf Fig.2:} The histograms of the mass distributions of our
clusters;
the upper panel shows the ACO-normalized distribution, while the lower
panel shows the EDCC-normalized distribution.

\end{document}